\documentclass[english]{webofc}
\usepackage[varg]{txfonts}

\woctitle{MENU 2013 -- 13th International Conference on Meson-Nucleon Physics and the Structure of the Nucleon}

\begin{document}
\title{Pion photo- and electroproduction in relativistic baryon ChPT}

\author{Lothar Tiator\inst{1}\fnsep\thanks{\email{tiator@kph.uni-mainz.de}} \and
        Stefan Scherer\inst{1} \and
        Marius Hilt\inst{1}
}

\institute{Institut f\"ur Kernphysik, Johannes Gutenberg-Universit\"at Mainz, D-55099 Mainz}

\abstract{We present a calculation of pion photo- and electroproduction in manifestly
Lorentz-invariant baryon chiral perturbation theory up to and including order $q^4$.
   We fix the low-energy constants by fitting experimental data in all available reaction channels.
   Our results can be accessed via a web interface, the so-called chiral MAID.
  }
\maketitle
\section{Introduction}
\label{sec:introduction}  In the middle of the 1980s, renewed interest in neutral pion
photoproduction at threshold was triggered by experimental data from Saclay and
Mainz
, which indicated a serious disagreement with the
predictions for the $s$-wave electric dipole amplitude $E_{0+}$ based on current algebra and PCAC.
This discrepancy was explained with the aid of ChPT
. Pion loops, which are
beyond the current-algebra framework, generate infrared singularities in the scattering amplitude
which then modify the predicted low-energy expansion of $E_{0+}$. Subsequently, several experiments
investigated pion photo- and electroproduction in the threshold region were performed at Mainz,
MIT-Bates, Saskatoon and TRIUMF, and on the theoretical side, all of the different reaction
channels of pion photo- and electroproduction near threshold were extensively investigated by
Bernard, Kaiser and Mei{\ss}ner within the framework of heavy-baryon chiral perturbation theory
(HBChPT). For a complete list of references, see Ref.~\cite{Hilt:2013fda}.

In the beginning, the manifestly Lorentz-invariant or relativistic formulation of ChPT (RChPT) was
abandoned, as it seemingly had a problem with respect to power counting when loops containing
internal nucleon lines come into play. Therefore, HBChPT became a standard tool for the analysis of
pion photo- and electroproduction in the threshold region. In the meantime, the development of the
infrared regularization (IR) scheme \cite{Becher:1999he} and the extended on-mass-shell (EOMS)
scheme \cite{Gegelia:1999gf} offered a solution to the power-counting problem, and RChPT became
popular again.

We present a calculation of pion photo- and electroproduction on the nucleon in manifestly
Lorentz-invariant baryon chiral perturbation theory up to and including chiral order $p^4$. Within
this framework we analyze $\pi^0$ and charged photo- and electroproduction data in the threshold
region, most of them obtained at the Mainz Microtron, MAMI. We also compare our results with the
dynamical model DMT~\cite{Kamalov:2001qg} and the unitary and causal effective field theory of
Gasparyan and Lutz ~\cite{Gasparyan:2010xz}.

\section{Pion photo- and electroproduction}
\label{sec:electroproduction}

For pion photoproduction with polarized photons from an unpolarized target without recoil
polarization detection, the cross section can be written in the following way with the unpolarized
cross section $\sigma_0$ und the photon beam asymmetry $\Sigma$.
\begin{equation}
\frac{d \sigma}{d \Omega} =  \sigma_0 \left( 1 - P_T \Sigma \cos 2 \varphi \right).
\end{equation}
For $\pi^0$ photoproduction on the proton, both observables are very precisely measured in the
threshold region, allowing an almost model independent partial wave
analysis~\cite{Hornidge:2012ca}.

For pion electroproduction, in the one-photon-exchange approximation, the differential cross
section can be written as
\begin{equation}
\frac{d\sigma}{d\mathcal{E}_fd\Omega_fd\Omega_\pi^\textnormal{\,cm}}
=\Gamma\frac{d\sigma_v}{d\Omega_\pi^\textnormal{\,cm}},
\end{equation}
where $\Gamma$ is the virtual photon flux and $d\sigma_v/d\Omega_\pi^\textnormal{\,cm}$ is the pion
production cross section for virtual photons.

\begin{figure}[htbp]
    \centering
        \includegraphics[width=0.70\textwidth]{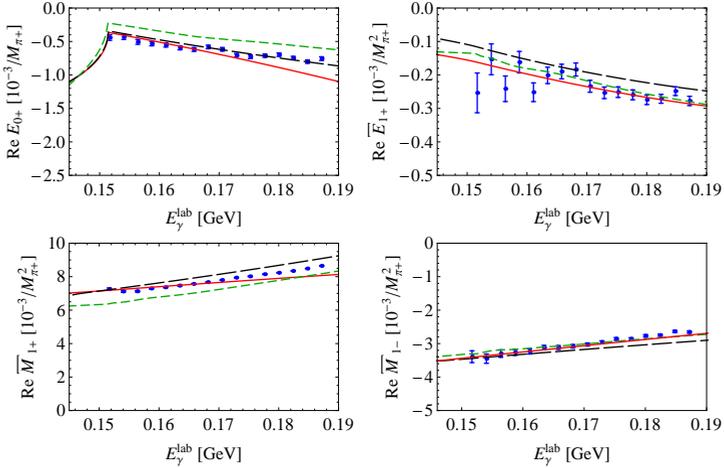}
    \caption{(Color online) $S$- and reduced $P$-wave multipoles for $\gamma+p\rightarrow p+\pi^0$.
    The solid (red) curves show our RChPT calculations at $O(q^4)$.
    The short-dashed (green) and long-dashed (black) curves are
    the predictions of the DMT model \cite{Kamalov:2001qg} and the GL model
    \cite{Gasparyan:2010xz}, respectively.
    The data are from Ref.\ \cite{Hornidge:2012ca}.}
    \label{fig:multipolesexportphyspi0p}
\end{figure}

For an unpolarized target and without recoil polarization detection, the virtual-photon
differential cross section for pion production can be further decomposed as
\begin{equation}
\frac{d\sigma_v}{d\Omega_\pi}=\frac{d\sigma_T}{d\Omega_\pi} +\epsilon\frac{d\sigma_L}{d\Omega_\pi}
+\sqrt{2\epsilon(1+\epsilon)}\frac{d\sigma_{LT}}{d\Omega_\pi}\cos\Phi_\pi
+\epsilon\frac{d\sigma_{TT}}{d\Omega_\pi}\cos{2\Phi_\pi}
+h\sqrt{2\epsilon(1-\epsilon)}\frac{d\sigma_{LT'}}{d\Omega_\pi}\sin{\Phi_\pi}, \label{eqn:wqparts}
\end{equation}
where it is understood that the variables of the individual virtual-photon cross sections
$d\sigma_T/d\Omega_\pi$ etc.~refer to the cm frame. For further details, especially concerning
polarization observables, see~\cite{Drechsel:1992pn}.

\begin{figure}[htbp]
    \centering
        \includegraphics[width=0.70\textwidth]{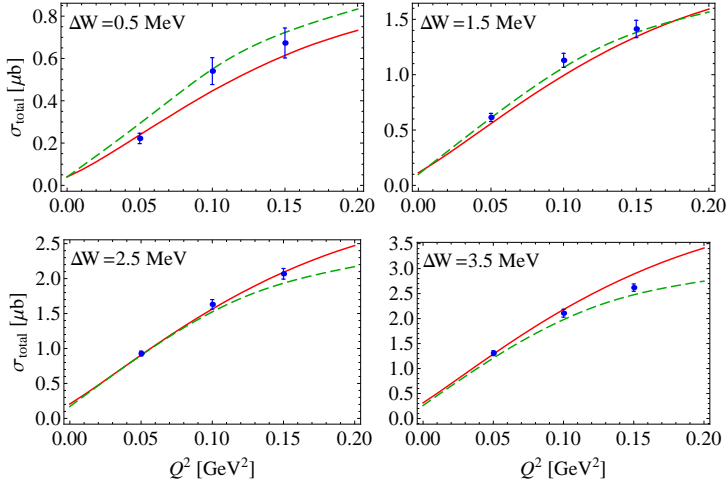}
    \caption{(Color online) Total cross sections in $\mu$b as a function of $Q^2$ for different cm energies
above threshold $\Delta W$ in MeV. The data are from Refs.\ \cite{Merkel:20092011}.}
    \label{fig:totcspi0electro}
\end{figure}

\section{Results and Conclusions}
\label{sec:results}

First, we show the real parts of the $S$ and $P$ waves in Fig.\ \ref{fig:multipolesexportphyspi0p}
together with single-energy fits of Ref.\ \cite{Hornidge:2012ca}. For comparison, we also show the
predictions of the Dubna-Mainz-Taipei (DMT) model \cite{Kamalov:2001qg} and the covariant, unitary,
chiral approach of Gasparyan and Lutz (GL) \cite{Gasparyan:2010xz}. The multipole $E_{0+}$ agrees
nicely with the data in the fitted energy range. The reduced $P$ waves
$\overline{E}_{1+}=E_{1+}/q_\pi$ and $\overline{M}_{1-}=M_{1-}/q_\pi$ with the pion momentum
$q_\pi$ in the c.m. frame agree for even higher energies with the single energy fits. The largest
deviation can be seen in $\overline{M}_{1+}$. This multipole is related to the $\Delta$ resonance
and the rising of the data above 170 MeV can be traced back to the influence of this resonance. As
we did not include the $\Delta$ explicitly, this calculation is not able to fully describe its
impact on the multipole.

\begin{figure}[htbp]
    \centering
        \includegraphics[width=0.70\textwidth,clip=true]{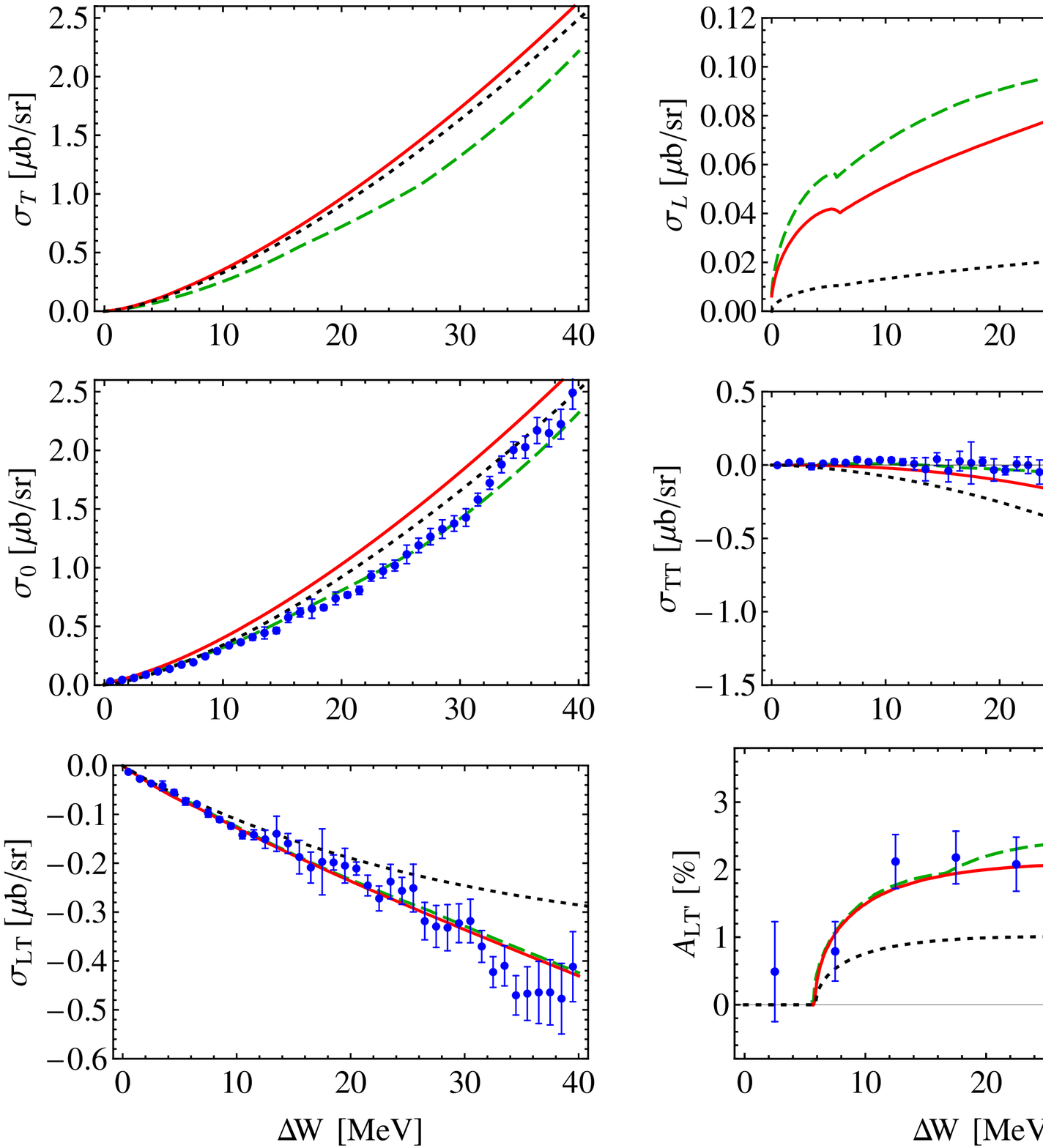}
    \caption{(Color online) Coincidence cross sections $\sigma_0$, $\sigma_{TT}$, and $\sigma_{LT}$
    in $\mu$b/sr and beam asymmetry
    $A_{LT'}$ in $\%$ at constant $Q^2=0.05$~GeV$^2$,
    $\Theta_\pi=90^\circ$, $\Phi_\pi=90^\circ$, and $\epsilon=0.93$  as a function of
    $\Delta W$ above threshold.
    The solid (red) lines show our RChPT calculations at $O(q^4)$ and the dotted (black) lines are the
    heavy-baryon ChPT calculations of Ref.~\cite{Bernard:1996ti}.
    The dashed (green) curves are obtained from the DMT
    model \cite{Kamalov:2001qg}.
    The data are from Ref.\ \cite{Weis:2007kf}.}
    \label{fig:weispi0electro}
\end{figure}

For electroproduction, in Fig.\ \ref{fig:totcspi0electro} we show the total cross section
$\sigma_{total}=\sigma_T+\epsilon\sigma_L$ in the threshold region together with the experimental
data \cite{Merkel:20092011}, and in Fig.~\ref{fig:weispi0electro} we compare our results for the
coincidence cross sections $\sigma_0$, $\sigma_{TT}$, $\sigma_{LT}$ and the beam asymmetry
$A_{LT'}$ with the experimental data of Ref.\ \cite{Weis:2007kf} and the results of HBChPT
\cite{Bernard:1996ti} and the DMT model \cite{Kamalov:2001qg}. In general, the DMT model gives a
very good description of all observables and amplitudes in the threshold region and can be used as
a guideline for theoretical calculations in cases where experimental data do not exist. The HBChPT
calculations shown in Fig.~\ref{fig:weispi0electro} were fitted to these data and are taken from
Ref.~\cite{Weis:2007kf}. In contrast, our RChPT calculation is not fitted to these data, as all
LECs were already determined with other data before. While HBChPT gives a better description for
the unpolarized cross section $\sigma_0(\Theta_\pi)=\sigma_T(\Theta_\pi)+\epsilon
\sigma_L(\Theta_\pi)$ than our RChPT calculation, a comparison with the separated cross sections
$\sigma_T$ and $\sigma_L$ shows that this is mainly due to a longitudinal cross section which is
much too small in the HBChPT fit. For the other observables $\sigma_{LT}$, $\sigma_{TT}$, and the
asymmetry $A_{LT'}$, RChPT compares much better to the data than HBChPT. It is interesting to note
that the asymmetry $A_{LT'}$ depends only weakly on LECs and has an important contribution from the
parameter-free pion loop contribution.

For $\Theta_\pi=\Phi_\pi=90^\circ$ and for $\epsilon\approx 1$ we find in very good approximation
the simple form
\begin{equation}
A_{LT'}(90^\circ) \approx
\frac{\sqrt{2\epsilon(1-\epsilon)}\;\sqrt{Q^2/k_0^2}\;(-P_2)\;\mbox{Im}(L_{0+})}{P_3^2}\,,
\end{equation}
where $P_2=3E_{1+}-M_{1+}+M_{1-}$ and $P_3=2M_{1+}+M_{1-}$. Therefore, this asymmetry is very
sensitive to the imaginary part of the longitudinal $S$ wave $L_{0+}$, hence practically
independent of LECs. This is very similar to the case of the target asymmetry $T$ for $\gamma p\to
p\pi^0$ which we discussed in Ref.~\cite{Hilt:2013uf}. There, the target asymmetry is shown to be
the ideal polarization observable to measure $\mbox{Im}(E_{0+})$.

In summary we have shown for the first time a chiral perturbation theory approach that consistently
can describe all pion photo- and electroproduction processes in the threshold region equally well.
Our relativistic chiral perturbation theory calculation is also online available within the MAID
project as ChiralMaid under http://www.kph.uni-mainz.de/MAID/.

\begin{acknowledgement}
This work was supported by the Deutsche Forschungsgemeinschaft (SFB 443 and 1044).
\end{acknowledgement}

\end{document}